\documentclass[10pt]{article} 

\usepackage{felix}

\newcommand{\st}{\ensuremath{\mbox{s.t.}}}

\title{\textbf{On the Complexity of Nucleolus Computation for Bipartite \texorpdfstring{$b$}{b}-Matching Games}
  \thanks{We acknowledge the support of the Natural Sciences and Engineering Research Council of Canada (NSERC).
  Cette recherche a \'et\'e financ\'ee par le Conseil de recherches en sciences naturelles et en g\'enie du Canada (CRSNG).}}

\date{\today}

\author{Jochen K\"onemann \\
  University of Waterloo \\
  Waterloo, Ontario N2L 1A2 \\
  {\it jochen@uwaterloo.ca}
  \and Justin Toth \\
  University of Waterloo \\
  Waterloo, Ontario N2L 1A2 \\
  {\it wjtoth@uwaterloo.ca}
  \and Felix Zhou
  \thanks{Present Address:
  Yale University,
  New Haven, CT 06511,
  {\it felix.zhou@yale.edu}} \\
  University of Waterloo \\
  Waterloo, Ontario N2L 1A2 \\
  {\it cfzhou@uwaterloo.ca}
}

\begin{document}

\maketitle

\begin{abstract}\noindent
  We explore the complexity of nucleolus computation in $b$-matching games on bipartite graphs.
  We show that computing the nucleolus of a simple $b$-matching game is \NP-hard when $b\equiv 3$
  even on bipartite graphs of maximum degree 7.
  We complement this with partial positive results in the special case where $b$ values are bounded by $2$.
  In particular, we describe an efficient algorithm when a constant number of vertices satisfy $b_v = 2$
  as well as an efficient algorithm for computing the non-simple $b$-matching nucleolus when $b\equiv 2$.
\end{abstract}

\section{Introduction}

\paragraph{}
Consider a network of companies such that any pair with a pre-existing
business relationship can enter into a deal that generates revenue, and at any given time every company has the capacity to fulfill a limited number of deals.
This is an example of a scenario that can be modeled as a cooperative $b$-matching game. 

A \emph{cooperative game} is a pair $(N,\nu)$ where $N$ is a finite set of \emph{players} and $\nu: 2^N \rightarrow \R$ is a \emph{value function} which maps subsets of players, known as \emph{coalitions} to a total value that their cooperation would generate.
In the special case of \emph{simple cooperative $b$-matching games},
we are given an underlying graph $G=(N, E)$, vertex values
$b:N\to \Z_+$, and edge weights $w:E\to \R$.  The set of {\em players}
in the game corresponds to the vertices $N$, and $w(uv)$ denotes the
value {earned} when $u, v\in N$ collaborate. 
For a coalition $S \subseteq N$, $\nu(S)$ corresponds to the maximum weight of a $b$-matching in $G[S]$ using each edge at most once. More formally, $\nu(S)$ is the optimal value of 
\begin{align*}
  &\max w(M) \\
  \abs{M \cap \delta(v)} &\leq b_v &\forall v \in S \\
  M &\subseteq E[S].
\end{align*}
On the other hand, in a \emph{non-simple cooperative $b$-matching game},
$\nu(S)$ is modified to allow $M$ to be a multiset
but we still require the the underlying set to be a subset of $E[S]$.

A central question in cooperative game theory is to study how the total revenue generated through the cooperation of the players is shared amongst the players themselves. An \emph{allocation} $x \in \R^N$ is a vector whose entries indicate the value each player should receive. 
Not all allocations are equally desirable.
Cooperative game theory gives us the language to model desirable allocations which capture notions such as fairness and stability.

An allocation $x \in \R^N$ is called \emph{efficient} if its entries sum to
$\nu(N)$; i.e., if $\sum_{i \in N}x_i := x(N) = \nu(N)$. Efficiency stipulates that an allocation should distribute the total value generated by the \emph{grand coalition} $N$.
We say $x$ is an \emph{imputation} if it is efficient and satisfies
\emph{individual rationality}: $x(i) \geq \nu(\set{i})$ for all $i$ in $N$.
Individual rationality captures the notion that each player should be assigned at least the value they can earn on their own.
In a $b$-matching game, $\nu(\{i\}) = 0$ and individual rationality simplifies to non-negativity.

The natural extension of individual rationality would be coalitional rationality, i.e.\ stipulating that for any coalition $S$, $x(S) \geq \nu(S)$. Allocations which satisfy such a property are said to lie in the \emph{core} of the game. Core allocations can be considered highly stable in the sense that no subset of players can earn more value by deviating from the grand coalition.

The core is well-known to be non-empty in {\em bipartite} $b$-matching games \cite{deng1999},
but may be empty in general matching games.
It is in fact known that the core of a matching game instance is non-empty
if and only if the {\em fractional matching linear program} is integral \cite{deng1999}.
For example, the core of the matching instance given by an odd-cycle with unit weights is empty.

Since the core may be empty, we need a more robust solution concept. Given an allocation, we let $e(S, x) := x(S) - \nu(S)$ be the
\emph{excess} of coalition $S\sset N$ with respect to
$x$. Informally, the excess measures the {\em satisfaction}
of coalition $S$: the higher the excess of $S$, the more satisfied its
players will be. We can rephrase the core as the set of all imputations where all coalitions have non-negative excess.

% We call a $b$-matching {\em simple} if each edge of our graph is
% included at most once; call a $b$-matching {\em non-simple}
% otherwise. Hereinafter, we will assume all $b$-matchings are simple
% unless otherwise specified. 

Instead of requiring all excesses to be non-negative, we can maximize the excess of the worst off coalitions.
Consider the following linear program
\begin{align*}
  &\max \epsilon &&(P_1) \\
  \st ~~ x(N) &= N \\
  x(S) &\geq \nu(S) + \epsilon &&\forall S \subset N \\
  x(i) &\geq \nu(\set{i}) &&\forall i\in N
\end{align*}
and let $\epsilon_1$ be its optimal solution.
The \emph{least core} is the set of allocations $x$ such that $(x, \epsilon_1)$ is optimal for ($P_1$). The least core is always non-empty.

For $b$-matching games when the core is non-empty, the least core coincides with the core. When the core is empty, the least core tries to maximize the satisfaction of the coalitions who are worst off in the game.
The least core, and by extension the core, both suffer from the fact that they are not in general unique.
Furthermore, the least core does nothing to improve the satisfaction of coalitions which are not the worst off.
This motivates the definition of the nucleolus, first introduced by Schmeidler \cite{schmeidler1969}.

For an allocation $x$, we write $\theta(x)\in \R^{2^N-2}$
as the vector whose entries are $e(S, x)$ for all $\varnothing\neq S\subsetneq N$
sorted in non-decreasing order.
This is a listing of the satisfactions of coalitions from worst off to best off.
The \emph{nucleolus} is defined as the allocation which lexicographically maximizes $\theta(x)$ over the imputation set. 
In a sense, the nucleolus is the most stable allocation.
In Schmeidler's paper introducing the nucleolus, the author proved, among other things, that it is unique.

We now have sufficient terminology to state our main result, proven in \Cref{sec:hardness}.

\begin{thm}\label{thm:nucleolus hard}
  The problem of deciding whether an allocation is equal to the nucleolus
  of an unweighted bipartite 3-matching game is \NP-hard,
  even in graphs of maximum degree 7.
\end{thm}

Kern and Paulusma posed the question of computing the nucleolus
for general matching games as an open problem in 2003 \cite{kern2003}.
In 2008, Deng and Fang conjectured this problem to be \NP-hard \cite{deng2008}.
This problem has been reaffirmed to be of interest in 2018 \cite{biro2018}.
In 2020, K\"onemann, Pashkovich, and Toth
proved the nucleolus of weighted matching games to be polynomial time computable \cite{konemann2020}.

On one hand, computing the nucleolus of unweighted $b$-matching games
when $b\geq 3$ is known to be \NP-hard for general graphs \cite{biro2019}.
However, the gadget graph in their hardness proof has many odd cycles.
On the other hand, Bateni et al. provided an efficient algorithm to compute the nucleolus in bipartite $b$-matching games
when one side of the bipartition is restricted to $b_v = 1$ and the other side is unrestricted \cite{bateni2010}.
Thus it is a natural question whether the nucleolus of bipartite $b$-matching games
is polynomial-time computable.
\Cref{thm:nucleolus hard} answers this question in the negative.

The basis of this result is a hardness proof for \emph{core} separation in unweighted bipartite 3-matching games \cite{biro2018}.
However, extending this to a hardness proof of nucleolus computation requires significant technical innovation.
Towards this end, we introduce a new problem in \Cref{sec:2csg}, a variant of the cubic subgraph problem, and prove that it is \NP-hard.
Then, in \Cref{sec:nucleolus hard}, we reduce the decision variant of nucleolus computation to our new problem,
which yields the result.

In \Cref{sec:positive}, we complement \Cref{thm:nucleolus hard} with efficient algorithms to compute the nucleolus
in two relevant cases when $b\leq 2$.
\Cref{sec:constant 2} explores the scenario when only a constant number of vertices satisfy $b_v = 2$
and \Cref{sec:non-simple} delves into the case when we relax the constraints to allow for non-simple $b$-matchings.

\begin{thm}\label{thm:b<=2}
  Let $G$ be a simple bipartite graph with bipartition $N = A\dot\cup B$
  and $k\geq 0$ a universal constant independent of $\card N$.
  Let $b\leq 2$ be some node-incidence capacity.
  \begin{enumerate}[(i)]
    \item Suppose $b_v = 2$ for all $v\in A$ but $b_v = 2$ for at most $k$ vertices of $B$,
      then the nucleolus of the simple weighted $b$-matching game on $G$ can be computed in polynomial time.
    \item If $b\equiv 2$,
      then the nucleolus of the non-simple weighted $b$-matching game on $G$ can be computed in polynomial time.
  \end{enumerate}
\end{thm}
We emphasize that for simple $b$-matching games,
$\nu(\cdot)$ is defined by $b$-matchings where each edge is picked at most once.
For non-simple $b$-matching games,
we allow each edge to be picked multiple times.

\subsection{Related Work}\label{sec:related}
\paragraph{}
The \emph{assignment game}, introduced by Shapley and Shubik \cite{shapley1971}, is the special case of simple $b$-matching games where $b$ is the all ones vector and the underlying graph is bipartite.
This was generalized to \emph{matching games} for general graphs by Deng, Ibaraki, and Nagamochi \cite{deng1999}.
Solymosi and Raghavan \cite{solymosi1994algorithm} showed how to compute the nucleolus in an unweighted assignment game. Kern and Paulusma \cite{kern2003} later gave a nucleolus computation algorithm for all unweighted matching games. Paulusma \cite{paulusma2001complexity} extended this result to all node-weighted matching games.
An application of assignment games is towards cooperative procurement from the field of supply chain management \cite{drechsel2010}.

The nucleolus is surprisingly ancient,
appearing as far back in history as a scheme for bankruptcy division in the Babylonian Talmud \cite{aumann1985}.
Modern research interest in the nucleolus is not only based on its widespread application \cite{branzei2005,lemaire1984},
but also the complexity of computing the nucleolus,
which seems to straddle the boundary between \P and \NP.

In a similar fashion to how we will define $b$-matching games,
a wide variety of combinatorial optimization games can be defined \cite{deng1999}.
Here the underlying structure of the game is based on the optimal solution
to some underlying combinatorial optimization problem.
One might conjecture that the complexity of nucleolus computation for a combinatorial optimization game
lies in the same class as its underlying combinatorial optimization problem.
However, this is not in general true.
For instance, nucleolus computation is known to be \NP-hard for network flow games \cite{deng2009},
weighted threshold games \cite{elkind2007},
and spanning tree games \cite{faigle1998,faigle2000}.
On the other hand, polynomial time algorithms are known for computing the nucleolus in special cases of network flow games \cite{deng2009,potters2006},
directed acyclic graph games \cite{solymosi2016,sziklai2017},
spanning tree games \cite{granot1996,kuipers2000},
$b$-matching games \cite{bateni2010},
fractional matching games \cite{faigle2001},
weighted voting games \cite{elkind2009}, convex games \cite{faigle2001}, and dynamic programming games \cite{konemann2020general}.

One possible application of cooperative matching games is to network bargaining \cite{willer1999, easley2012}.
In this setting, a population of players are connected through an underlying social network.
Each player engages in a profitable partnership with at most one of its neighbours and the profit must be shared between the participating players in some equitable fashion.
Cook and Yamagishi \cite{cook1992} proposed a profit-sharing model that generalizes Nash's famous 2-player bargaining solution \cite{nash1950} as well as validates empirical findings from the lab setting.

Both the pre-kernel and least-core are solution concepts which contain the nucleolus.
It is well-known that the pre-kernel of a cooperative game may be non-convex and even disconnected \cite{kopelowitz1967, stearns1968}.
Nonetheless, Faigle, Kern, and Kuipers showed how to compute a point in the intersection of the pre-kernel and least-core in polynomial time given a polynomial time oracle to compute the minimum excess coalition for a given allocation \cite{faigle2001}.
The authors later refined their result to compute a point within the intersection of the core and lexicographic kernel \cite{faigle2006},
a set which also contains the nucleolus.

The complexity of computing the nucleolus of $b$-matching games remained open for bipartite graphs,
and for b-matching games where $b\leq 2$.
In \Cref{thm:nucleolus hard},
we show that the former is indeed \NP-hard to compute
and give an efficient algorithm for a special case of the latter in \Cref{sec:positive}.
Some positive results about the core were proved for non-bipartite graphs in \cite{biro2018}.

\subsection{The Kopelowitz Scheme}
A more computational definition of the nucleolus is provided by the Kopelowitz Scheme \cite{kopelowitz1967}.
Recall the linear program $(P_1)$
and let $\epsilon_1$ be its optimal value.
Write $\mcal S := 2^N\setminus \set{\varnothing, N}$
to denote the set of all non-trivial coalitions.
Finally, put
\[
  \mcal S_1
  := \set{S\in \mcal S:
  \text{$x(S) = \nu(S) + \epsilon_1$ for all optimal solutions $(x, \epsilon_1)$}}.
\]
We say $\mcal S_1$ are the coalitions which are \emph{fixed} in $(P_1)$.

For $\ell\geq 2$, let $(P_\ell)$ be the linear program
\begin{align}
  &\max \epsilon &&(P_\ell) \label{eq:kopelowitz-scheme} \\
  x(N) &= \nu(N) \\
  x(S) &= \nu(S) + \epsilon_i &&\forall i\leq \ell-1, \forall S\in \mcal S_i \\
  x(S) &\geq \nu(N) + \epsilon &&\forall S\in \mcal S\setminus \left( \bigcup_{i=1}^{\ell-1} \mcal S_i \right)
\end{align}
Recursively, we set
\[
  \mcal S_\ell
  := \set{S\in \mcal S:
  \forall i\leq \ell-1, S\notin \mcal S_i,
  \text{$x(S) = \nu(S) + \epsilon_\ell$ for all optimal solutions $(x, \epsilon_\ell)$ to $(P_\ell)$}}.
\]
These are the coalitions which are fixed in $(P_\ell)$
but not in any $(P_i)$ for $i\leq \ell-1$.
This hierarchy of linear programs terminates when the dimension of the feasible region becomes $0$,
at which point the unique feasible solution is the nucleolus \cite{davis1965}.

Directly solving each $(P_\ell)$ requires solving a linear program with an exponential number of constraints in terms of the number of players and hence takes exponential time with respect to the input \footnote{Cooperative games we are interested in have a compact representation roughly on the order of the number of players. For example $b$-matching games can be specified by a graph, $b$-values and edge weights rather than explicitly writing out $\nu$.}.
Moreover, the best general bound on the number of linear programs we must solve until we obtain a unique solution is the naive exponential bound $O(2^{|N|})$.
However, we are still able to use the Kopelowitz Scheme as a way to characterize the nucleolus in the proof of \Cref{thm:nucleolus hard}.

One way of solving exponentially sized linear programs is to utilize the polynomial time equivalence of optimization and separation \cite{khachiyan1979}.
That is, to develop a separation oracle and employ the ellipsoid method.
For our positive results, we will take this route.

Indeed, we will develop a polynomial-size formulation of each $(P_\ell)$
by pruning unnecessary constraints.
Not only does this enable us to solve each $(P_\ell)$ in polynomial time,
but we also reduce the number of iterations to a polynomial of the input size
since at least one inequality constraint is changed to an equality constraint per iteration.

It is of interest to consider a variation of the Kopelowitz Scheme by Maschler \cite{maschler1979}.
In this variation, the author defines $\mcal S_\ell$ as
\[
  \mcal S_\ell
  := \set{S\in \mcal S:
  \forall i\leq \ell-1, S\notin \mcal S_i,
  \text{$\exists c_S\in \R$ such that $x(S) = c_S$ for all optimal solutions $x$ of $(P_\ell)$}}.
\]
This way, at least 1 equality constraint is added to $(P_{\ell+1})$
which is linearly independent of all equality constraints in $(P_{\ell})$.
Hence the feasible region decreases by at least 1 dimension per iteration
and there are at most $\card N$ iterations before termination.

\section{Hardness}\label{sec:hardness}
We consider $b$-matching games for $b\equiv 3$ and uniform weights.
The goal of the this section is to prove \Cref{thm:nucleolus hard}.

The idea of the proof is inspired the hardness proof of core separation employed in \cite{biro2018}
and the hardness proof in \cite{deng2009}.
We reduce the problem into a variation of \textsc{Cubic Subgraph} which is \NP-hard \cite{plesnik1984}
through a careful analysis of several iterations of the Kopelowitz Scheme.
However, it is not clear that our variation of \textsc{Cubic Subgraph} is \NP-hard
and we significantly extend the techniques from \cite{plesnik1984} to prove its hardness.

\begin{prob}[\textsc{Two from Cubic Subgraph}]
  Let $G$ be an arbitrary graph.
  Decide if $G$ contains a subgraph $H$
  (not necessarily induced)
  such that there are $u\neq v\in V(H)$ satisfying
  \[
    \deg_H(w) = 
    \begin{cases}
      2, &w\in \set{u, v} \\
      3, &\text{else}
    \end{cases}
  \]
  for all $w\in V(H)$.
  We say that $H$ is a Two from Cubic Subgraph.
\end{prob}

\begin{thm}\label{thm:2csg hardness}
  \textsc{Two from Cubic Subgraph} is $\NP$-hard,
  even in bipartite graphs of maximum degree 4.
\end{thm}

This theorem is proven in \Cref{sec:2csg}
and is crucial to our proof of \Cref{thm:nucleolus hard} in \Cref{sec:nucleolus hard}.

\subsection{The Reduction}\label{sec:nucleolus hard}
\begin{figure}
  \centering
  \includegraphics[scale=0.6]{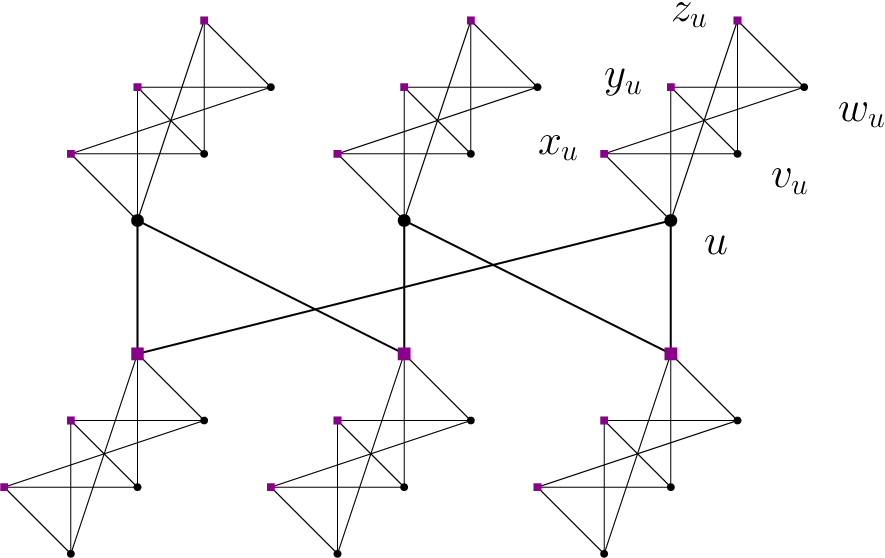}
  \caption{The gadget graph from \cite{biro2018}.}
  \label{fig:gadget}
\end{figure}

Hereinafter, $G = (N, E)$ is a bipartite instance of \textsc{Two from Cubic Subgraph}.
We assume that $E\neq \varnothing$ so that $\card N\geq 2$.
Take $G^* := (N^*, E^*)$ to be the following bipartite gadget graph depicted in \Cref{fig:gadget},
initially proposed in \cite{biro2018}:
For each original vertex $u\in N$,
create 5 new vertices $v_u, w_u, x_u, y_u, z_u$.
Then, define $N^* := N\cup \set{v_u, w_u, x_u, y_u, z_u: u\in N}$.
To obtain $E^*$ from $E$,
we add edges until $(\set{u, v_u, w_u}, \set{x_u, y_u, z_u})$ is a $K_{3, 3}$ subgraph for every $u\in N$.
Notice that $E^*\setminus E$ forms a maximum 3-matching in $G^*$
with cardinality $\frac32 \card{N^*}$.

In \Cref{fig:gadget}, the bigger vertices with bolded edges indicate the original graph
and the smaller vertices with thinner edges were added to obtain the gadget graph.
The square and circular vertices depict a bipartition of $G^*$.
Observe that the maximum degree of $G^*$ is the maximum degree of $G$ plus 3.

For each $u\in N$, define
$T_u := \set{v_u, w_u, x_u, y_u, z_u}$
as well as $V_u := T_u\cup \set{u}$.
We say that $T_u$ are the \emph{gadget vertices} of $u$
and $V_u$ is the \emph{complete gadget} of $u$.
Remark that each $V_u$ induces a connected component within the maximum 3-matching $E^*\setminus E$ on $G^*$.

Let $\Gamma = (N^*, \nu)$ be the unweighted 3-matching game on $G^*$.

In \Cref{lem:uniform nucleolus},
we show that if no two from cubic subgraph of $G$ exists,
then the nucleolus is precisely $x^* \equiv \frac32$.
Conversely, we prove in \Cref{lem:not uniform nucleolus}
that the existence of a two from cubic subgraph
implies that $x^*$ cannot be the nucleolus.
The proof of \Cref{thm:nucleolus hard} follows from the above lemmas
and the hardness of \textsc{Two from Cubic Subgraph} proven in \Cref{sec:2csg}.
Remark that the degree bound follows from the degree bound in \Cref{thm:2csg hardness}
and the fact that our gadget graph increases the maximum degree of the original graph by 3.

\begin{lem}\label{lem:core allocations}
  Let $M$ be a maximum 3-matching in $G^*$.
  Let $\mcal C$ be the set of connected components of $G^*[M]$.
  Then for all core allocations $x$ and every component $C\in \mcal C$,
  $x(C) = \nu(C)$.
\end{lem}

\begin{pf}
  Observe that
  \[
    x(N^*) \geq x(M) = \sum_{C\in \mcal C} x(C) \geq \sum_{C\in \mcal C} \nu(C) = \sum_{C \in \mcal C} |M\cap E(C)| = |M| = \nu(N^*)
  \]
  with the first inequality following from the fact that $x\geq 0$
  and the second inequality following from the assumption that $x$ is in the core.
\end{pf}

\begin{lem}\label{lem:uniform nucleolus}
  If $G$ does not contain a two from cubic subgraph,
  the uniform allocation $x^*\equiv \frac32$ is the nucleolus of $\Gamma$.
\end{lem}

\begin{table}
  \centering
  \tabulinesep=1.2mm
  \begin{tabu}{|c|c|c|r|} 
    \hline
    $\card{S\cap \set{u, v_u, w_u}}$ & $\card{S\cap \set{x_u, y_u, z_u}}$ & $\nu(S)$ & $e(S, x^*)$ \\
    \hline\hline
    0 & 1 & 0 & $\frac32$ \\
    0 & 2 & 0 & $3$ \\
    0 & 3 & 0 & $\frac92$ \\
    1 & 0 & 0 & $\frac32$ \\
    1 & 1 & 1 & $2$ \\
    1 & 2 & 2 & $\frac52$ \\
    1 & 3 & 3 & $3$ \\
    2 & 0 & 0 & $3$ \\
    2 & 1 & 2 & $\frac52$ \\
    2 & 2 & 4 & $2$ \\
    2 & 3 & 6 & $\frac32$ \\
    3 & 0 & 0 & $\frac92$ \\
    3 & 1 & 3 & $3$ \\
    3 & 2 & 6 & $\frac32$ \\
    \hline
  \end{tabu}
  \caption{The excess computation for \Cref{lem:uniform nucleolus} when $x^*\equiv \frac32, S\subsetneq V_u$.}
  \label{tab:uniform nucleolus}
\end{table}

\begin{pf}
  We argue using the Kopelowitz Scheme.
  Put $(P_k)$ as the $k$-th LP in the Kopelowitz Scheme.

  We check through computation that for all $u\in N$ and $S\subsetneq V_u$,
  $e(S, x^*) \geq \frac32$.
  See \Cref{tab:uniform nucleolus}.

  Let $\epsilon_1$ be the optimal objective value to $(P_1)$.
  We claim that $\epsilon_1 = 0$.
  By core non-emptiness, we have $\epsilon_1 \geq 0$.
  Since $E\neq \varnothing$, we can choose $u\in N$ so that $V_u\subsetneq N$.
  But $V_u$ is a connected component of the maximum 3-matching $E^*\setminus E$.
  Thus by \Cref{lem:core allocations},
  $V_u$ is a coalition for which $e(V_u, x) = 0$
  for all core allocations $x$.
  It follows that $\epsilon_1 = 0$ and the set of optimal solutions to $(P_1)$ is precisely the core.

  We now claim that
  \begin{equation}
    \mcal S_1 = \set*{S\sset N^*: S = \bigcup_{u\in S\cap N} V_u}.
    \label{eq:S1}
  \end{equation}
  These are the unions of complete gadgets.

  Let $x$ be an optimal solution to $(P_1)$ (core allocation).
  Clearly, if $S$ is a union of complete gadgets,
  then $e(S, x^*) = 0$ due to \Cref{lem:core allocations}.
  This shows the reverse inclusion in \Cref{eq:S1}.
  Notice that $\nu(S) \leq \frac32 \card S = x^*(S)$ by the definition of a 3-matching,
  so $x^*$ is an optimal solution to $(P_1)$
  and we may assume that $x = x^*$.

  We claim that
  \begin{equation}
    \forall S\notin \mcal S_1, e(S, x^*) \geq \frac32.
    \label{eq:uniform allocation excess lower bound}
  \end{equation}
  This shows that if $S$ is not a union of complete gadgets,
  then there is some optimal solution of $(P_1)$ for which $S$ is not fixed in $(P_1)$
  and hence $S\notin \mcal S_1$.
  Thus the inclusion in \Cref{eq:S1} would hold.
  
  \Cref{eq:uniform allocation excess lower bound} is true if $S = \set{u}$ for some $u\in N^*$.
  If $S\sset N$ with $\card S\geq 2$, then $\nu(S) \leq \frac32 \card{S} - 2$.
  Otherwise, the edges of a maximum 3-matching in $G[S]$ induce a two from cubic subgraph.
  Thus $e(S, x^*) \geq \frac32\card S - \left( \frac32\card S - 2 \right) \geq 2$.

  It remains to consider the case when there is some $u\in N$ such that $T_u\cap S\neq \varnothing$.
  The argument here is similar to the one employed in \cite{biro2018}:
  First suppose $S\cap V_u = V_u$
  and remark that there is a maximum 3-matching in $G^*[S]$
  where all edges adjacent to $u$ are contained in the edge set induced by $V_u$.
  Thus deleting such edges yields a 3-matching in $G^*[S\setminus V_u]$.
  \begin{align*}
    &e(S\setminus V_u, x^*) \\
    &\leq e(S, x^*) - x^*(V_u) + \card{E^*(V_u)} &&\nu(S\setminus V_u) \geq \nu(S) - \card{E^*(V_u)} \\
    &= e(S, x^*) - 9 + 9 \\
    &= e(S, x^*).
  \end{align*}
  We can remove as many of these complete gadgets from $S$ as possible to obtain some coalition $S'$
  with $e(S', x^*) \leq e(S, x^*)$.
  It suffices then to lower bound $e(S', x^*)$.

  If $S' = \varnothing$, then $S\in \mcal S_1$ by definition.
  In addition, if $S'\sset N$, there is again nothing to prove.
  Thus there must be some $u'\in S'$ such that $\card{T_{u'}\cap S'} \geq 1$
  and $S'\cap V_{u'} \neq V_{u'}$.

  Suppose now that $\card{S'\cap T_{u'}} \leq 4$.
  Remark that if $M$ is a 3-matching in $G^*[S']$,
  then $M\setminus E^*(S'\cap T_{u'} \cup \set{u'})$ is a 3-matching in $G^*[S'\setminus T_{u'}]$.
  \begin{align*}
    &e(S'\setminus T_{u'}, x^*) \\
    &\leq e(S', x^*) - x^*(S'\cap T_{u'}) + \card{E^*(S'\cap T_{u'}\cup \set{u'})} &&\nu(S'\setminus T_{u'}) \geq \nu(S') - \card{E^*(S'\cap T_{u'}\cup \set{u'})} \\
    &= e(S', x^*) - \frac32\card{S'\cap T_{u'}} + \card{E^*(S'\cap T_{u'}\cup \set{u'})} \\
    &\leq e(S', x^*).
  \end{align*}
  Finally, if $\card{S'\cap T_{u'}} = 5$, we are required to have $u'\notin S'$.
  If $M$ is a 3-matching in $G^*[S']$,
  then $M\setminus E^*(S'\cap T_{u'})$ is a 3-matching in $G^*[S'\setminus T_{u'}]$.
  \begin{align*}
    &e(S'\setminus T_{u'}, x^*) \\
    &\leq e(S', x^*) - x^*(S'\cap T_{u'}) + \nu(T_{u'}) &&\nu(S'\setminus T_{u'}) \geq \nu(S') - \nu(T_{u'}) \\
    &= e(S', x^*) - 5\cdot \frac32 + 6 \\
    &\leq e(S', x^*).
  \end{align*}
  We may thus again repeatedly remove vertices of $N^*\setminus N$ until we arrive back at the base case of $S'\sset N$.

  So \Cref{eq:uniform allocation excess lower bound} holds
  as all other coalitions have strictly greater excess with respect to $x^*$.

  We now argue that $\epsilon_2 = \frac32$.
  Observe $\nu(N^*) = \frac32\card{N^*}$ implies that $\min_{a\in N^*} x(a) \leq \frac32$
  for any allocation $x$ in the core
  and thus also for feasible solutions to $(P_2)$
  as well as the nucleolus.
  It follows that $\epsilon_2 \leq \frac32$.
  But \Cref{eq:uniform allocation excess lower bound} shows that this upper bound is attained by $x^*$.

  For all feasible solutions $x$ to $(P_2)$,
  $x(a)\geq \frac32$ for all $a\in N^*$.
  But we cannot have some $x(a) > \frac32$,
  or else $x(N^*) > \frac32 \card{N^*}$ and $x$ would not be an allocation.
  Since the singleton coalitions are fixed in $(P_2)$,
  it must be that $x^*\equiv \frac32$ is the nucleolus.
\end{pf}

\begin{lem}\label{lem:not uniform nucleolus}
  If $G$ contains a two from cubic subgraph,
  then the nucleolus of the gadget graph is not $x^*\equiv \frac32$.
\end{lem}

\begin{table}
  \centering
  \tabulinesep=1.2mm
  \begin{tabu}{|c|c|c|c|r|} 
    \hline
    $\card{S\cap \set{u}}$ & $\card{S\cap \set{v_u, w_u}}$ & $\card{S\cap \set{x_u, y_u, z_u}}$ & $\nu(S)$ & $e(S, x^*)$ \\
    \hline\hline
    0 & 0 & 1 & 0 & $\frac32-\frac\delta5$ \\
    0 & 0 & 2 & 0 & $3-\frac{2\delta}5$ \\
    0 & 0 & 3 & 0 & $\frac92-\frac{3\delta}5$ \\
    0 & 1 & 0 & 0 & $\frac32-\frac\delta5$ \\
    0 & 1 & 1 & 1 & $2-\frac{2\delta}5$ \\
    0 & 1 & 2 & 2 & $\frac52-\frac{3\delta}5$ \\
    0 & 1 & 3 & 3 & $3-\frac{4\delta}5$ \\
    0 & 2 & 0 & 0 & $3-\frac{2\delta}5$ \\
    0 & 2 & 1 & 2 & $\frac52-\frac{2\delta}5$ \\
    0 & 2 & 2 & 4 & $2-\frac{4\delta}5$ \\
    0 & 2 & 3 & 6 & $\frac32-\delta$ \\
    1 & 0 & 0 & 0 & $\frac32+\delta$ \\
    1 & 0 & 1 & 1 & $2+\frac{4\delta}5$ \\
    1 & 0 & 2 & 2 & $\frac52+\frac{3\delta}5$ \\
    1 & 0 & 3 & 3 & $3+\frac{2\delta}5$ \\
    1 & 1 & 0 & 0 & $3+\frac{4\delta}5$ \\
    1 & 1 & 1 & 2 & $\frac52+\frac{3\delta}5$ \\
    1 & 1 & 2 & 4 & $2+\frac{2\delta}5$ \\
    1 & 1 & 3 & 6 & $\frac32+\frac\delta5$ \\
    1 & 2 & 0 & 0 & $\frac92+\frac{3\delta}5$ \\
    1 & 2 & 1 & 3 & $3+\frac{2\delta}5$ \\
    1 & 2 & 2 & 6 & $\frac32+\frac\delta5$ \\
    \hline
  \end{tabu}
  \caption{The excess computation for \Cref{lem:not uniform nucleolus} when $x_\delta, S\subsetneq V_u$.}
  \label{tab:not uniform nucleolus}
\end{table}

\begin{pf}
  Suppose that $G$ contains a two from cubic subgraph.
  We will show that $x^*\equiv \frac32$ is not an optimal solution to $(P_2)$.
  Recall that the nucleolus is necessarily an optimal solution to each LP in the Kopelowitz Scheme.
  This would thus yield the desired result.

  Let us introduce a parameter as follows:
  \begin{equation*}
    \Delta :=
    \begin{cases}
      0, &\text{$G$ contains a cubic subgraph} \\
      1, &\text{$G$ contains a two from cubic subgraph but no cubic subgraphs}
    \end{cases}
    \label{eq:tilted allocation delta}
  \end{equation*}

  Let $N'\sset N$ be the vertices in the cubic subgraph or
  the vertices of the two from cubic subgraph if no cubic subgraph exists.
  Then
  \begin{equation}
    e\left( N', x^* \right) = \frac32\card{N'} - \left( \frac32\card{N'} - \Delta \right) = \Delta.
    \label{eq:tilted allocation delta excess}
  \end{equation}
  In particular, the minimum excess over all coalitions in $\mcal S$ is at most $\Delta$.

  For $0 < \delta < \frac12$, define
  \begin{equation*}
    x_\delta(a) :=
    \begin{cases}
      \frac32 + \delta, &a\in N \\
      \frac32 - \frac\delta5, &a\in N^*\setminus N
    \end{cases}
    \label{eq:tilted allocation}
  \end{equation*}
  We check by computation in \Cref{tab:not uniform nucleolus} that
  \begin{equation}
    \forall u\in N, \forall S\subsetneq V_u, e(S, x_\delta) \geq \frac32 - \delta
    \label{eq:tilted allocation complete gadget lower bound}
  \end{equation}
  The coalitions with minimum excess among such coalitions is $S = T_u$ for some $u\in N$.

  We claim that $\epsilon_1 = 0$ and
  \begin{equation*}
    \mcal S_1 = \set*{S\sset N^*: S = \bigcup_{u\in S\cap N} V_u}.
  \end{equation*}
  is again the union of complete gadgets.
  The fact that $\epsilon_1 = 0$ is clear from our previous lemma.
  Moreover, it is clear from our prior work that the unions of complete gadgets must be fixed in $(P_1)$.
  We need only show that $e(S, x_\delta) > 0$ if $S$ is not a union of complete gadgets.
  This would show that if $S$ is not a union of complete gadgets,
  then there is some allocation (in particular $x_\delta$) for which $S$ is not fixed in $(P_1)$.

  If $S = \set{a}$ for some $a\in N^*$, then $e(S, x_\delta) \geq \frac32 - \frac\delta5 > 0$.

  When $S\sset N$, we have
  \begin{align*}
    e(S, x_\delta)
    &= x_\delta(S) - \nu(S) \\
    &\geq \left( \frac32 + \delta \right)\card S - \left( \frac32\card S - \Delta \right) \\
    &= \delta\card S + \Delta \\
    &> 0.
  \end{align*}

  Suppose now that there is some $u\in N$ such that $S\cap T_u\neq \varnothing$.
  Once again, if $S\cap V_u = V_u$, $e(S\setminus V_u, x_\delta) \leq e(S, x_\delta) - 9 + 9 = e(S, x_\delta)$.
  We can thus remove all complete gadgets from $S$ to obtain another coalition $S'$.
  If $S' = \varnothing$, then $S\in \mcal S_1$.
  Similar to before, if $S'\sset N$, we are back at the base case.

  We proceed assuming there is some $u'\in S'$ such that $\card{S'\cap T_{u'}}\geq 1$.
  Observe that $\nu(S') \leq \nu(S'\setminus T_{u'}) + \nu(S'\cap V_{u'})$.
  This is because any maximum 3-matching on $S'$ is a disjoint union of 3-matchings on $S'\setminus T_{u'}$ and $S'\cap V_{u'}$.
  
  Suppose $u'\in S'$.
  We must have $\card{S'\cap T_{u'}}\leq 4$.
  \begin{align*}
    e(S', x_\delta)
    &= x_\delta(S'\setminus T_{u'}) + x_\delta(S'\cap V_{u'}) - x_\delta(u') - \nu(S') \\
    &\geq x_\delta(S'\setminus T_{u'}) - \nu(S'\setminus T_{u'})
    + \left[ x_\delta(S'\cap V_{u'}) - x_\delta(u') \right] - \nu(S'\cap V_{u'}) \\
    &\geq e(S'\setminus T_{u'}, x_\delta)
    + \card{S\cap T_{u'}}\left( \frac32 - \frac\delta5 \right) - \card{E^*(S'\cap V_{u'})} \\
    &\geq e(S'\setminus T_{u'}, x_\delta) - \frac45\delta.
  \end{align*}
  Suppose now that $u'\notin S'$.
  In this case,
  \begin{align*}
    e(S', x_\delta)
    &= x_\delta(S'\setminus T_{u'}) + x_\delta(S'\cap T_{u'}) - \nu(S') \\
    &\geq x_\delta(S'\setminus T_{u'}) - \nu(S'\setminus T_{u'}) + x_\delta(S'\cap T_{u'}) - \nu(S'\cap T_{u'}) \\
    &= e(S'\setminus T_{u'}, x_\delta) + e(S'\cap T_{u'}, x_\delta) \\
    &\geq e(S'\setminus T_{u'}, x_\delta) + \left( \frac32 - \delta \right) &&\text{by \Cref{eq:tilted allocation complete gadget lower bound}}
  \end{align*}

  By repeatedly removing vertices of $N^*\setminus N$,
  we see that
  \begin{align*}
    e(S', x_\delta)
    &\geq e(S'\cap N, x_\delta)
    + \sum_{u\in S'\cap N: S'\cap T_u\neq \varnothing} \left[ - \frac45\delta \right]
    + \sum_{u\in N\setminus S': S'\cap T_u\neq \varnothing} \left[ \frac32 - \delta \right] \\
    &\geq \delta\card{S'\cap N}
    + \Delta - \card{S'\cap N} \left( \frac45 \delta \right)
    + \card{\set{u\in N\setminus S': S'\cap T_u\neq \varnothing}} \left( \frac32 - \delta \right) \\
    &= \frac\delta5 \card{S'\cap N} + \Delta
    + \card{\set{u\in N\setminus S': S'\cap T_u\neq \varnothing}} \left( \frac32 - \delta \right) \\
    &\geq \frac\delta5 + \Delta.
  \end{align*}
  The last inequality follows from the assumption that $S'\neq \varnothing$.
  In particular, at least one of $S'\cap N$
  or $\set{u\in N\setminus S': S'\cap T_u\neq \varnothing}$ is non-empty.
  This shows that $\epsilon_1 = 0$ is indeed the optimal solution to $(P_1)$.
  Moreover, $\mcal S_1$ is again the union of complete gadgets.
  
  As an immediate corollary to the proof,
  $\epsilon_2 \geq \frac\delta5 + \Delta > \Delta$.
  Recall there was a coalition $N'\sset N$ satisfying \Cref{eq:tilted allocation delta excess}.
  It follows that $x^* \equiv \frac32$ is not an optimal solution to $(P_2)$
  and therefore cannot be the nucleolus.
\end{pf}

\subsection{Two from Cubic Subgraph}\label{sec:2csg}
In this subsection, we prove that \textsc{Two from Cubic Subgraph},
from which we reduce to nucleolus testing, is \NP-hard.

\begin{prob}[Exact Cover by 3-Sets]
  Let a ground set $X = \set{a_1, a_2, \dots, a_{3k}}$
  and a collection of 3-element subsets of $X$
  $S = \set{S_1, S_2, \dots, S_t}$ be given.

  Decide if there is a subcollection $Y\sset S$
  where each element $a_i\in X$ is included in exactly one subset $S_j\in Y$.
\end{prob}
It is well known that \textsc{Exact Cover by 3-Sets} is \NP-hard,
even if every element of the ground set belongs to exactly three subsets \cite{gonzalez1985}.
We reduce \textsc{Exact Cover by 3-Sets} to \textsc{Two from Cubic Subgraph}.

The idea is to construct two parallel instances of the bipartite gadget graph from \cite{plesnik1984}.
We will then focus on any ``non-trivial'' two from cubic subgraphs.

\subsubsection{Step I}
Let $X, S$ be an instance of the \textsc{Exact Cover by 3-Sets}
where every element of the ground set belongs to exactly three subsets.
Create the bipartite graph $G_0$ with bipartition $X\dot\cup S$,
where $a_iS_j\in E(G_0) \iff a_i\in S_j$.
The problem is reformulated as follows:
Does there exist a subgraph with vertex set $X\cup S'$
such that every vertex of $X$ has degree 1
and each vertex of $S'$ has degree 3?
Notice we require the entire ground set to be included in the subgraph vertex set.

\subsubsection{Step II}
Add $7k$ new vertices to $G_0$,
$b_1, b_2, \dots, b_{7k}$,
as follows.
Each $b_i, 1\leq i\leq 3k$ is adjacent to $a_i$ and $b_{3k+i}$.
If $i > 1$, then $b_i$ is also adjacent to $b_{3k+i-1}$.
Otherwise, $i=1$ and $b_i$ is also adjacent to $b_{6k}$.
Finally, each $b_{6k+j}, 1\leq j\leq k$,
is adjacent to $b_{3k+3j-2}, b_{3k+3j-1}, b_{3k+3j}$.

\begin{figure}
  \centering
  \includegraphics[scale=0.5]{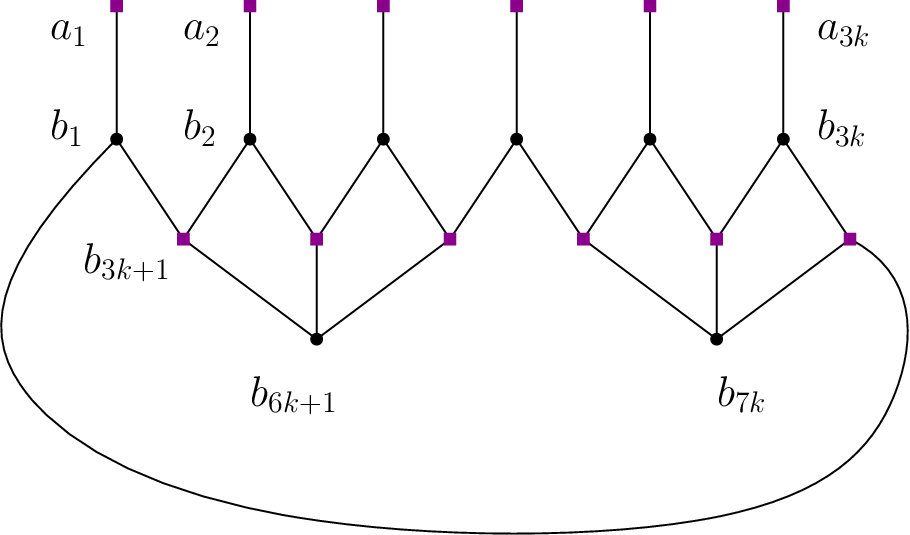}
  \caption{A subgraph of $G_1$ for $k=2$, depicting the changes in Step I.}
  \label{fig:G1}
\end{figure}

See \Cref{fig:G1}.
The square and circular vertices depict a bipartition of $G_1$.

Define the following tripartition of the newly added vertices
$B_1 := \set{b_i: 1\leq i\leq 3k}$,
$B_2 := \set{b_{3k+i}: 1\leq i\leq 3k}$,
and $B_3 := \set{b_{6k+i}: 1\leq i\leq k}$.

Let us refer to this graph as $G_1$.

\subsubsection{Step III}
At this step, we diverge from the work in \cite{plesnik1984}.
In Plesnik's hardness proof of \textsc{Cubic Subgraph},
the author substituted a complete bipartite graph at each $a_i$
so the resulting graph has a cubic subgraph
if and only if it has the desired subgraph in Step I.
We proceed by using a grid-like substitution
at each $a_i$ into two copies of $G_1$.

Let $G_2$ be the graph obtained after the following substitution:
At each vertex $a_i$,
we substitute the following gadget.

For $j=1,2,3$, Let $S_{i, j}\in S$ be the 3-element subsets containing $a_i$.
Add vertices $u_{i, j}, w_{i, j}, c_{u_{i, j}}, c_{w_{i, j}}$ for $j=1, 2, 3$.
For $j=1, 2, 3$, create the edges
\[
  w_{i, j-1}u_{i, j}, u_{i, j}w_{i, j}, w_{i, j}S_{i, j},
  u_{i, j}c_{u_{i, j}}, w_{i, j}c_{w_{i, j}}, c_{u_{i, j}}c_{w_{i, j}}.
\]
Here we understand $w_{i, 0} = b_i$.

Next, create a copy $G_2'$ of $G_2$.
For each $v\in G_2$, we let $v'$ denote the corresponding copy in $G_2'$.
Also, add the edges $c_{w_{i, j}}c_{w_{i, j}}', c_{u_{i, j}}c_{u_{i, j}}'$ for $1\leq i\leq 3k, 1\leq j\leq 3$.

\begin{figure}
  \centering
  \includegraphics[scale=0.5]{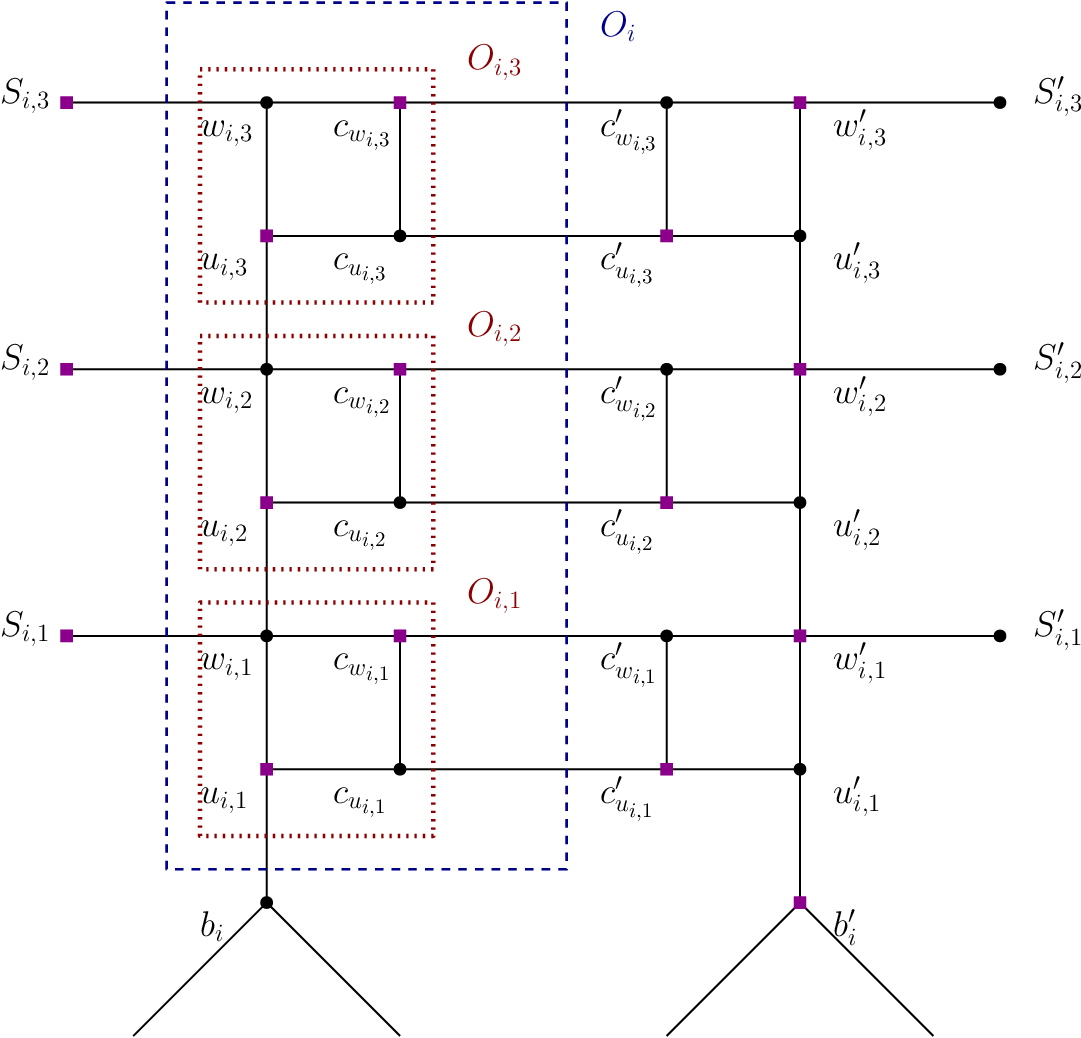}
  \caption{The substitution at what used to be $a_i, a_i'$.}
  \label{fig:Gmerge}
\end{figure}

See \Cref{fig:Gmerge}.
The square and circular vertices depict a bipartition of $G$.

Define $O_{i, j} := \set{u_{i, j}, w_{i, j}, c_{u_{i, j}}, c_{w_{i, j}}}$.
Let us call the vertices $O_i := \bigcup_{j=1}^3 O_{i, j}$
the \emph{ore} of $b_i\in B_1$ and similarly for $b_j'\in B_1'$.

Let this new graph be $G$.
In summary we have the vertex sets $B := B_3\cup B_2\cup B_1, S$ and $O_i, 1\leq i\leq 3k$
coming from the substitution in $G_1$.
Similarly $B', S', O_i', 1\leq i\leq 3k$ are the vertex sets from $G_1'$.
Let $A = B\cup S\cup \bigcup_{1\leq i\leq 3k} O_i$
and $A' = B'\cup S'\cup \bigcup_{1\leq i\leq 3k} O_i'$
be the vertices which originated from $G_1, G_1'$ respectively,
so $V(G) = A\dot\cup A'$.

\subsubsection{Locally Regular Graphs}

Before we can prove correctness we need to introduce the concept of locally regular graphs which our proof will heavily rely on.

Let $G$ be a graph and $v\in V(G)$.
Here we introduce a common definition $N(v) := \set{w\in V(G): vw\in E(G)}$
referred to as the \emph{neighbours} of $v$ in $G$.
We denote the number of neighbours of $v$ in $G$ by $\deg_G(v) := \card{N(v)}$.

\begin{defn}[Locally Regular]
  Let $G$ be an arbitrary graph and put $\varnothing\neq V'\sset V$.
  Suppose $G$ contains a subgraph $H$ such that for all $v\in V'\cap V(H)$,
  $\deg_H(v) = r$
  where $r\geq 1$ is a constant.

  We say that $H$ is a locally $(V', r)$-regular subgraph of $G$.
\end{defn}

Next, \Cref{lem:propagation} essentially shows the following:
Given a locally $(V', r)$-regular subgraph $H$ of a graph $G$
where $G[V']$ is highly vertex-connected
and there only a few vertices $v\in V'$ for which $\deg_G(v)\neq r$,
then $V(H)$ either contains all of $V'$ or is disjoint from $V'$

Before stating the lemma, let us introduce a notation.
For any graph $G$ and $V'\sset V(G)$, put $\Delta_{V'} := \max_{v\in V'} \deg_G(v)$.

\begin{lem}[Propagation]\label{lem:propagation}
  Let $G$ be an arbitrary graph and $\varnothing\neq V'\sset V(G)$
  be such that $G[V']$ is $\kappa$-vertex-connected
  for some $\kappa\geq 1$.

  Let $H$ be a locally $(V', r)$-regular subgraph of $G$
  for some $r\geq 1$.

  Put $V_r := \set{v\in V': \deg_G(v) = r}$.
  Then the following hold:
  \begin{enumerate}[(i)]
    \item If
      \begin{equation}
        \kappa - \card{V'\setminus V_r} \geq 1
        \label{eq:propagation assume1}
      \end{equation}
      and $V_r\cap V(H)\neq \varnothing$, then $V'\sset V(H)$.
    \item If
      \begin{equation}
        \kappa - \card{V'\setminus V_r} \geq \max(\Delta_{V'} - r, 1)
        \label{eq:propagation assume2}
      \end{equation}
      then $V'\not\sset V(H)$ implies $V'\cap V(H) = \varnothing$.
  \end{enumerate}
\end{lem}

\begin{figure}
  \centering
  \includegraphics[scale=0.5]{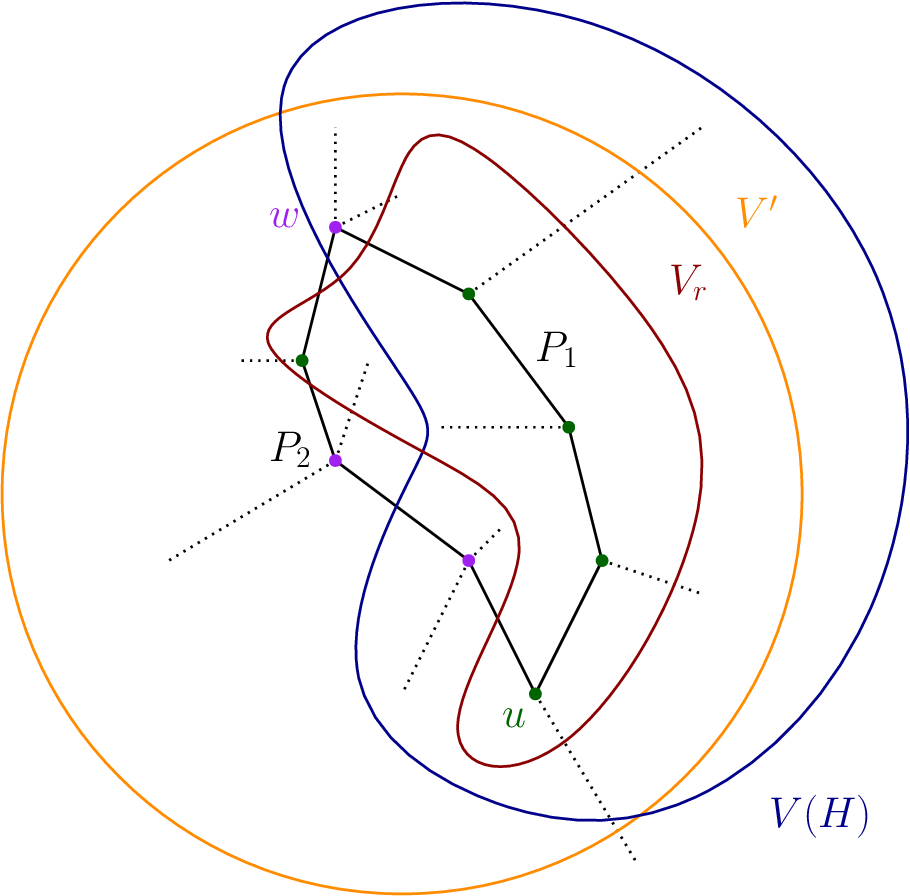}
  \caption{An illustration of \Cref{lem:propagation} with $r=3, \kappa=2$ and $\Delta_{V'} = 4$.}
  \label{fig:propagation}
\end{figure}

See \Cref{fig:propagation}.
The green vertices reside in $V_r$ while the purple vertices do not.
Notice how $P_2$ potentially leaves $H$ but $V(P_1)\sset V(H)$ as all internal vertices have degree $r=3$.

\begin{pf}[i]
  Let $G$ be an arbitrary graph and $\varnothing\neq V'\sset V(G)$
  be such that $G[V']$ is $\kappa$-vertex-connected for some $\kappa\geq 1$.
  Let $H$ be a locally $(V', r)$-regular subgraph of $G$ for some $r\geq 1$.
  
  Suppose now that $V_r\cap V(H)\neq \varnothing$
  and $\kappa - \card{V'\setminus V_r} \geq 1$.
  Pick $v\in V_r\cap V(H)$ and fix any $v\neq w\in V'$.
  The goal is to show that $w\in V(H)$.
  This implies $V'\sset V(H)$ by the arbitrary choice of $w$.
  
  Remark that $v, w\in V'$.
  Hence by Menger's theorem, there are $\kappa$ internally vertex-disjoint $vw$-paths in $G[V']$
  $P_1, P_2, \dots P_\kappa$.
  If one of these paths contains no internal vertices,
  then $w\in N(v)$ and the definition of local regularity 
  yields $w\in V(H)$.
  We proceed assuming all paths have internal vertices.

  Hence, by the assumption that $\kappa - \card{V'\setminus V_r}\geq 1$
  and the pigeonhole principle,
  at least one of these paths, say $P_1$,
  is composed internally only of vertices from $V_r$.
  Write $P_1: u, v_1, v_2, \dots, v_\ell, w$.
  Thus $\deg_G(v_i) = r$ for all $1\leq i\leq \ell$.

  Since $u\in V'\cap V(H)$,
  we can apply the definition of local regularity to see that $v_1\in V(H)$.
  Repeat this argument to see that $v_2, \dots, v_\ell\in V(H)$.
  Finally, since $w\in N(v_\ell)$,
  apply the definition of local regularity once more to conclude that $w\in V(H)$.

  This concludes the proof by our initial remark.
\end{pf}

\begin{pf}[ii]
  Suppose now that $\kappa - \card{V'\setminus V_r}\geq \max(\Delta_{V'}-r, 1)$ holds
  and there is some vertex $u\in V'\setminus V(H)$.
  Let $u\neq w\in V'$.
  There are again $\kappa$ internally vertex-disjoint $uw$-paths in $G[V']$,
  say $P_1, P_2, \dots, P_\kappa$.

  \underline{Case I: $w\in V_r$}
  Since $\kappa - \card{V'\setminus V_r}\geq 1$,
  and $w\in V_r\cap V(H)\neq \varnothing$,
  an application of (i) yields $V'\sset V(H)$.
  This contradicts the existence of $u$.

  \underline{Case II: $w\in V'\setminus V_r$}
  There are at most $\card{V'\setminus V_r}-1$ vertices from $V'\setminus V_r$
  which can be internal vertices within our $\kappa$ paths.
  We will consider $u$ as an internal vertex.
  By the assumption that $\kappa - \card{V'\setminus V_r}\geq \Delta_{V'}-r$ and the pigeonhole principle,
  at least $\kappa-(\card{V'\setminus V_r} - 1)\geq \Delta_{V'} - r + 1$
  paths are composed internally of vertices only from $V_r$.

  Consider such a path $P_i: u, v_1, v_2, \dots, v_\ell, w$
  composed internally of vertices from $V_r$.
  Since $u\notin V(H)$, $v_1$ cannot attain degree $r$ in $H$ thus $v_1\notin V(H)$ as well.
  Repeating this argument, we see that $v_2, \dots, v_\ell\notin V(H)$.

  Since there are at least $\Delta_{V'}-r+1$ such paths,
  $\deg_H(w) \leq \Delta_{V'} - (\Delta_{V'} - r + 1) = r-1$.
  Thus the definition of local regularity asserts that $w\notin V(H)$.

  It follows that $V'\cap V(H) = \varnothing$ by the arbitrary choice of $w$ as desired.
\end{pf}

\subsubsection{Correctness}
Now we will show that $G$ contains a two from cubic subgraph
if and only if there is an exact cover of $X$ by 3-sets.

\begin{defn}[Trivial Two from Cubic Subgraph]
  Let $H$ be a two from cubic subgraph of $G$ with special vertices $u, v$.
  If $uv\in E(G)\setminus E(H)$,
  then $H+uv$ is a cubic subgraph of $G$
  and $H$ is a trivial two from cubic subgraph of $G$.
\end{defn}
Observe that $G$ has a cubic subgraph if and only if
it has a trivial two from cubic subgraph.

In \Cref{thm:csg hardness} we show that an exact cover exists
if and only if $G$ contains a (trivial two from) cubic subgraph.
This proves the forward direction as the existence of an exact cover
implies the existence of a two from cubic subgraph.
The converse is also simplified,
since if there is a trivial two from cubic subgraph,
it implies the existence of an exact cover by 3-sets.
We need only prove that if $G$ contains a non-trivial two from cubic subgraph,
then there is an exact cover.
This is done in \Cref{prop:nontrivial 2csg}.

First, we prove that an instance of \textsc{Exact Cover by 3-Sets} is a ``yes'' instance
if and only if the constructed gadget graph contains a (trivial two from) cubic subgraph.

For graphs $H, H'$,
we say $H'$ was obtained from $H$ by \emph{adding a path}
if $H'$ contains a path $P$ with endpoints $u\neq v$ such that
$V(H) = V(H')\setminus (V(P)\setminus \set{u, v})$
and $E(H) = E(H')\setminus E(P)$.

We will use the following elementary graph theory result.

\begin{prop}[Ear Decomposition; \cite{diestel2016}]\label{prop:ear decomp}
  A graph is 2-vertex-connected
  if and only if it can be constructed from a cycle
  by successively adding a path.
\end{prop}

\begin{lem}\label{lem:2-con}
  $G[B]$ and $G[O_{i, j}]$ for all $i, j$ are 2-vertex-connected.
\end{lem}

\begin{pf}
  \underline{$G[B]$ is 2-Vertex-Connected:}
  Let us argue by ear decomposition (\Cref{prop:ear decomp}).
  Notice first that $G[B_1\cup B_2]$ is a cycle.
  We successively add each vertex $b_{6k+j}\in B_3$ and its incident edges to obtain $G[B]$ as follows:
  First add the path consisting of vertices $b_{3k+3j-2}, b_{6k+j}, b_{3k+3j-1}$.
  Then add the path consisting of vertices $b_{6k+j}, b_{3k+3j}$.

  Thus $G[B]$ is indeed 2-vertex-connected.

  \underline{$G[O_{i, j}]$ is 2-Vertex-Connected:}
  $G[O_{i, j}]$ is simply a cycle and hence 2-vertex-connected.
\end{pf}

\begin{lem}\label{lem:B-control}
  Suppose $H$ is a locally $(B, 3)$-regular subgraph of $G$.
  Let $v\in B$ be arbitrary.

  Then $v\in V(H)$ implies $B\sset V(H)$
  and $v\notin V(H)$ implies $B\cap V(H) = \varnothing$.
  In particular, either $B\sset V(H)$ or $B\cap V(H) = \varnothing$.
\end{lem}

\begin{pf}
  Since $G[B]$ is 2-vertex-connected by \Cref{lem:2-con},
  we can simply apply \Cref{lem:propagation}
  with $V' = B, \kappa = 2, r = 3, \Delta_B = 3$.

  Note that $V'\setminus V_r = \varnothing$.
  Hence both assumptions from \Cref{lem:propagation}
  (\Cref{eq:propagation assume1}, \Cref{eq:propagation assume2})
  are satisfied.
\end{pf}

\begin{lem}\label{lem:O-control}
  Suppose $H$ is a locally $(O_{i, j}, 3)$-regular subgraph of $G$
  for some $i, j$ and fix some $v\in O_{i, j}$.

  Then $v\in H$ implies $O_{i, j}\sset V(H)$
  and $v\notin H$ implies $O_{i, j}\cap V(H) = \varnothing$.
  In particular, either $O_{i, j}\sset V(H)$ or $O_{i, j}\cap V(H) = \varnothing$.
\end{lem}

\begin{pf}
  Since $G[O_{i, j}]$ is 2-vertex-connected by \Cref{lem:2-con},
  we can simply apply \Cref{lem:propagation}
  with $V' = O_{i, j}, \kappa = 2, r = 3, \Delta_{O_{i, j}} = 4$.

  Observe that $\card{V'\setminus V_r} \leq 1$.
  Hence both assumptions from \Cref{lem:propagation}
  (\Cref{eq:propagation assume1}, \Cref{eq:propagation assume2})
  are satisfied.
\end{pf}

\begin{thm}\label{thm:csg hardness}
  There is an exact cover of $X$ by 3-sets
  if and only if $G$ contains a cubic subgraph.
\end{thm}

\begin{pf}
  \underline{$(\implies)$}
  Let $Y\sset S$ be the exact cover of $X$.
  
  We will describe the vertices from $A$ of a cubic subgraph
  and take the exact copy within $A'$.

  First, take all vertices $b\in B$.
  Fix $b_i\in B_1$.
  There is exactly one $S_{i, j}\in Y$ such that $a_i\in S_{i, j}$.
  Let $P_i$ be the unique $b_iS_{i, j}$-path in the ore of $b_i$.
  Put $Q_i := \bigcup_{j: O_{i, j}\cap V(P_i)\neq \varnothing} O_{i, j}$.
  Then we take $B\cup \left( \bigcup_i Q_i \right)\cup Y$
  as well as their copies in $A'$ to be the vertex set of our cubic subgraph.

  \underline{$(\impliedby)$}
  Let $H$ be a cubic subgraph of $G$.
  First, we claim that $B\sset V(H)$.

  Suppose otherwise.
  By \Cref{lem:B-control}, which states that if $H$ is locally $(B,3)$-regular then either $B \subseteq V(H)$ or $B\cap V(H) =\varnothing$, we see that $B\cap V(H) = \varnothing$.
  
  Fix some $u_{i, 1}$.
  \Cref{lem:2-con} states that $G[O_{i, 1}]$ is 2-vertex-connected.
  From our work above, we must have $u_{i, 1}\notin V(H)$.
  \Cref{lem:O-control} states that if $H$ is locally $(O_{i,j}, 3)$-regular then either $O_{i, j}\sset V(H)$ or $O_{i, j}\cap V(H) = \varnothing$. Since the former cannot happen, $O_{i, 1}\cap V(H) = \varnothing$.
  Thus $u_{i, 2}\notin V(H)$.

  Repeating this argument for $u_{i, 2}, u_{i, 3}$ shows that $V(H)\sset S$.
  But then $H$ cannot be 3-regular.
  By contradiction, $B\sset V(H)$.

  Finally, let us fix some $b_i\in B_1$ and remark that $b_i\in V(H)$ implies $u_{i, 1}\in V(H)$.
  Once more, apply \Cref{lem:O-control} to see that $O_{i, 1}\sset V(H)$.
  If $H$ contains the edge $w_{i, 1}S_{i, 1}$,
  then it contains no more vertices in the ore of $b_i$.
  Otherwise, we repeat this argument to see that $H$ also contains $O_{i, 2}$, etc.
  All in all, there is exactly one subset $S_j\in S$ which is adjacent to the ore of $b_i$.

  Put $Y := S\cap V(H)$ as the sub-collection.
  By our work above, $Y$ covers each element exactly once.
  Thus $Y$ is the desired exact cover.
\end{pf}

Now, we prove the converse:
If $G$ contains a two from cubic subgraph,
then there is an exact cover of $X$ by 3-sets.
The key insight is that the structure of our gadget forces the two inconvenient vertices
to be in different copies.
Then, we apply \Cref{lem:propagation} multiple times
to show that the arguments made in \Cref{thm:csg hardness} still follow
despite the existence of those two vertices.

\begin{defn}[Lacking]
  Let $H$ be a non-trivial two from cubic subgraph.
  We say $v\in V(H)$ is lacking if $\deg_H(v) = 2$.
\end{defn}
Notice that if $v$ is lacking with $vw\in E(G)\setminus E(H)$
and $\deg_G(w) = 3$, then it must be that $w\notin V(H)$
or else $H$ was not a non-trivial two from cubic subgraph.

\begin{prop}\label{prop:nontrivial 2csg}
  If $G$ contains a two from cubic subgraph,
  then there is an exact cover of $X$ by 3-sets.
\end{prop}

Recall that $A = B\cup S\cup \bigcup_{1\leq i\leq 3k} O_i$
and $A' = B'\cup S'\cup \bigcup_{1\leq i\leq 3k} O_i'$
are the vertices which originated from $G_1, G_1'$ respectively,
so $V(G) = A\dot\cup A'$.

\begin{pf}
  Let $H$ be a non-trivial two from cubic subgraph. 

  If $A\cap V(H)\neq \varnothing$ and no vertex of $A$ is lacking,
  then there is an exact cover. We prove this in \Cref{lem:easy x3c}.
  Thus it suffices to consider the cases where $A\cap V(H) = \varnothing$
  or some vertex of $A$ is lacking.
  
 We claim that $A\cap V(H)\neq \varnothing\neq A'\cap V(H)$. We prove this claim in \Cref{lem:both nonempty}.
  Moreover, if either $A, A'$ does not contain a lacking vertex,
  then there is an exact cover. We defer the proof to \Cref{lem:separation}.
  Thus we may as well assume that each of $A, A'$ contains a lacking vertex
  or we are done.
  It is also well-defined to say \emph{the} lacking vertex of $A$ or $A'$.

  We will show in \Cref{lem:B struc} and \Cref{lem:S struc} respectively that
  $B\sset V(H)$ and $S$ does not contain a lacking vertex.
  Thus with the lacking vertex of $A$ not in either $B, S$.
  It must thus reside in the ore of some $b_i$.

  For all other ores of $b_j, j\neq i$,
  \Cref{lem:O-control} forces that ore to be adjacent to exactly one vertex of $S$.
  But we must have $\card{\delta(S\cap V(H))} \equiv 0\mod 3$.
  The only way for this to hold is if the ore of $b_i$ is also adjacent to exactly one vertex of $S$.

  Thus despite the ore of $b_i$ containing a lacking vertex,
  $Y := S\cap V(H)$ covers every ground set element exactly once,
  concluding our proof.
\end{pf}

\begin{lem}\label{lem:easy x3c}
  Let $H$ be a non-trivial two from cubic subgraph. 
  Suppose $A\cap V(H)\neq \varnothing$
  and that no vertex of $A$ is lacking.
  
  Then there is an exact cover.
\end{lem}

\begin{pf}
  The proof of this claim is similar to the reverse direction for \Cref{thm:csg hardness}.

  First note that $A\cap V(H)\neq \varnothing$ implies that $B\sset V(H)$.
  Indeed, if any vertex $b\in B$ is not a vertex of $V(H)$,
  then the \Cref{lem:B-control} ensures that $B\cap V(H) = \varnothing$.
  But then repeated applications of \Cref{lem:O-control}
  show that $H$ does not contain any ore vertices.
  So $H$ contains no vertices of $A$, which is a contradiction.

  Since $B\sset V(H)$,
  let us focus on the ores.
  For every ore of some $b_i$,
  repeated application of \Cref{lem:O-control} force $H$ to choose exactly one edge of the form $w_{i, j}S_{i, j}$.

  Thus taking $Y := V(H)\cap S$ defines an exact cover of $X$.
\end{pf}

\begin{lem}\label{lem:both nonempty}
  Let $H$ be a non-trivial two from cubic subgraph. 

  Then $A\cap V(H)\neq \varnothing\neq A'\cap V(H)$.
\end{lem}

\begin{pf}
  By symmetry, it suffices to consider the case where $A'\cap V(H) = \varnothing$.

  Observe that $\card{B_1\cap V(H)} \leq 2$
  since if the ore of $b_i\in B_1$ does not contain a lacking vertex,
  then $c_{w_{i, j}}\notin V(H)$ for all $j=1, 2, 3$.
  Hence by \Cref{lem:O-control}, $O_{i, j}\cap V(H) = \varnothing$ for each $j=1, 2, 3$.
  The claim then follows.

  But then $\card{B_2\cap V(H)}\leq 1$
  since if $b_j\in B_2$ is not adjacent to both its neighbors in $B_1$,
  it cannot attain degree 3.
  This implies $B_3\cap V(H) = \varnothing$
  as no $b_i\in B_3$ attains degree 3.
  These restrictions show that in fact $B\cap V(H) = \varnothing$.

  If $H$ does not contain any ore vertices, it is the empty graph.
  
  Suppose $H$ partially contains exactly one ore of $b_i\in B_1$.
  Then at least one vertex of the ore must be lacking.
  But then it is not possible for a vertex of $S$ to be of degree 2 or 3 so $S\cap V(H)=\varnothing$.
  Specifically, $w_{i, 3}\notin V(H)$.
  By an inductive argument, it is not hard to see that $O_{i, 3}, O_{i, 2}, O_{i, 1}\cap V(H) = \varnothing$.
  But then $H$ is actually the empty graph.

  Finally, suppose $H$ partially contains exactly two ores of $b_i\neq b_j\in B_1$.
  Observe that it can partially contain at most two ores in total,
  as each of them consumes one lacking vertex.
  But then $H$ cannot contain any vertex of $S$ since any vertex of $S$ cannot achieve degree 3.
  Similar to the previous case, $H$ is actually the empty graph.
\end{pf}

\begin{lem}\label{lem:separation}
  Let $H$ be a non-trivial two from cubic subgraph. 

  If either $A, A'$ does not contain a lacking vertex,
  there is an exact cover of $X$ by 3-sets.
\end{lem}

\begin{pf}
  Suppose not.
  By \Cref{lem:both nonempty},
  each of $A, A'$ is not empty and one of them,
  say $A$, does not contain a lacking vertex.

  The result follows then from \Cref{lem:easy x3c}.
\end{pf}

\begin{lem}\label{lem:all/none}
  Let $H$ be a non-trivial two from cubic subgraph
  where both $A, A'$ contains a lacking vertex.

  Either $B\sset V(H)$ or $B\cap V(H) = \varnothing$.
\end{lem}

\begin{pf}
  By \Cref{lem:2-con}, $G[B]$ is 2-vertex-connected.

  Suppose now that the statement does not hold.
  There is a vertex $b\in B\cap V(H)$
  and another vertex $b^*\in B\setminus V(H)$.

  By 2-vertex-connectedness, there are two vertex disjoint $bb^*$-paths $P_1, P_2$ in $G[B]$.
  Since $b^*\notin V(H)$,
  there are two distinct edges $b_ib_j, b_kb_\ell$
  which are contained in $P_1, P_2$ respectively such that $b_i, b_k\in V(H), b_j, b_\ell\notin V(H)$.

  If $b = b_i = b_k$, then $\deg_H(b) = 1$ which is a contradiction.
  Otherwise, $b_i\neq b_k$ and we notice that $\deg_H(b_i), \deg_H(b_k)\leq 2$.
  But only one of the two can be lacking and we have a contradiction.
\end{pf}

\begin{lem}\label{lem:B struc}
  Let $H$ be a non-trivial two from cubic subgraph
  where both $A, A'$ contains a lacking vertex.

  $B\sset V(H)$ and does not contain a lacking vertex.
\end{lem}

\begin{pf}
  By \Cref{lem:all/none},
  either $B\sset V(H)$ or $B\cap V(H) = \varnothing$.

  \underline{$B\sset V(H)$:}
  Suppose $B\cap V(H) = \varnothing$.
  Repeated application of \Cref{lem:O-control} to each ore shows that
  if some ore $O_i$ does NOT contain the lacking vertex,
  then $O_i\cap V(H) = \varnothing$.
  In particular, at most one such $O_i$ intersects non-trivially with $V(H)$.

  But then $S\cap V(H) = \varnothing$
  as no vertex of $S$ can attain degree 3 within $H$.
  It follows that $w_{i, 3}\notin V(H)$.
  A similar argument to the proof of \Cref{lem:both nonempty} shows that $O_i\cap V(H) = \varnothing$ as well.

  This is a contradiction since we assumed that $V(H)\cap A \neq \varnothing$.
  Thus $B\sset V(H)$.
  We now show that no vertex of $B$ is lacking.

  \underline{$B$ contains no lacking vertices:}
  This is easy for $b\in B_3\cup B_2$.
  Indeed, any such lacking vertex implies one of its neighbors in $B$ is not in $H$.
  This would contradict the assumption that $B\sset V(H)$.
  Similarly, if $b\in B_1$ is not adjacent to one of its neighbors in $B$,
  then we cannot have $B\sset V(H)$.
  The only possibility is some $b_i\in B_1$ not adjacent to $u_{i, 1}$.

  But then by the repeated application of \Cref{lem:O-control},
  we see that $H$ does not contain any vertices from the ore of $b_i$.
  But \Cref{lem:O-control} forces $H$ to contain vertices from all other ores.
  Moreover, each such ore is adjacent to exactly one vertex in $S$.
  Thus $\card{\delta(S\cap V(H))}\equiv -1\mod 3$
  which is a contradiction as no vertex of $S$ is lacking.
\end{pf}

\begin{lem}\label{lem:S struc}
  Let $H$ be a non-trivial two from cubic subgraph
  where both $A, A'$ contains a lacking vertex.

  $S$ does not contain a lacking vertex.
\end{lem}

\begin{pf}
  Suppose otherwise,
  that some vertex of $S$ is lacking.

  By \Cref{lem:B struc}, $B\sset V(H)$.
  Similar to before,
  repeated application of \Cref{lem:O-control} shows that $A$ contains vertices from each ore,
  and every ore is adjacent to exactly one vertex in $S$.
  
  It follows that $\card{S\cap V(H)}\equiv 0\mod 3$.
  But since $S$ contains a lacking vertex,
  the above is impossible.
\end{pf}

In all cases, whether $H$ is a trivial or non-trivial two from cubic subgraph,
there is an exact cover.
Moreover, the degree bound from \Cref{thm:2csg hardness} holds since $G$ is bipartite with maximum degree 4.

\section{Positive Results}\label{sec:positive}
In the case of $b\leq 2$,
we explore several variants of $b$-matching games
for which the nucleolus can be efficiently computed.

First, we will state an important ingredient.

Let $\Gamma = (N, \nu)$ be a cooperative game.
For $\mscr S\sset \mcal S$ and $x\in \R^N$, write $\theta^{\mscr S}(x)\in \R^{\mscr S}$
to denote the restricted vector containing the excess values $e(S, x)$ for $S\in \mscr S$
in non-decreasing order of excess.

\begin{defn}[Characterization Set]
  Let $\mscr S\sset \mcal S$ be a subset of the non-trivial coalitions.

  We say $\mscr S$ is a characterization set for the nucleolus of the cooperative game $\Gamma = (N, \nu)$
  if the lexicographic maximizer of $\theta^{\mscr S}(x)$ is a singleton
  that lexicographically maximizes the unrestricted vector $\theta(x)$.
\end{defn}

Intuitively, for $S\in \mcal S\setminus \mscr S$,
we can drop the constraint corresponding to $S$
from the Kopelowitz Scheme when computing the nucleolus.

\begin{prop}\label{prop:char-coalitions}
  Let $\Gamma = (N, \nu)$ be a cooperative game
  with non-empty core.
  Suppose $\mscr S$ is a polynomial sized characterization set for the nucleolus of $\Gamma$.

  The nucleolus of $\Gamma$ is polynomial time computable.
\end{prop}

Let $\mscr S\sset \mcal S$ be a characterization set of the nucleolus of some game $\Gamma$.
Consider the following tweak of the $\ell$-th iteration of Kopelowitz Scheme $(P_\ell')$ (with optimal value $\epsilon_\ell'$)
where we only have constraints corresponding to coalitions in the characterization set $\mscr S$
instead of every coalition.
The sets $\mcal S_\ell$ are defined in symmetric fashion as the coalitions from $\mscr S$
which are fixed in $(P_\ell)$ but not at any prior $(P_i)$.

\begin{align}
  &\max \epsilon &&(P_\ell') \\
  x(S) &= \nu(S) - \epsilon_i' &&\forall 0\leq i< \ell, \forall S\in \mcal S_i \\
  x(S) - \nu(S) &\geq \epsilon &&\forall S\in \mscr S\setminus \bigcup_{i=0}^{\ell-1} \mcal S_i
\end{align}

\begin{pf}
  The tweaked Kopelowitz Scheme computes the lexicographic maximizer of $\theta^{\mscr S}$.
  Since $\mscr S$ is polynomially sized,
  each linear program in the scheme can be solved in polynomial time.
\end{pf}

We are now ready to state the theorem found in \cite{granot1998}.

\begin{thm}[\cite{granot1998}]\label{thm:char-coalitions}
  Let $\Gamma = (N, \nu)$ be a cooperative game with non-empty core.

  The non-empty collection $\mscr S\sset \mcal S$ is a characterization set for the nucleolus of $\Gamma$
  if for every $S\in \mcal S\setminus \mscr S$
  there exists a non-empty subcollection $\mscr S_S$ of $\mscr S$ such that
  \begin{enumerate}[(i)]
    \item For all $T\in \mscr S_S$ and core allocations $x$,
      $e(T, x) \leq e(S, x)$.
    \item There are scalars $\lambda_T\in \R$ such that the characteristic vector $\chi_S\in \set{0, 1}^N$ of $S$ satisfies
      $\chi_S = \sum_{T\in \mscr S_S\cup \set{N}} \lambda_T \chi_T$.
  \end{enumerate}
\end{thm}

\begin{cor}\label{cor:char-bmatching}
  Let $\Gamma = (N, \nu)$ be a not necessarily simple weighted $b$-matching game
  with non-empty core.
  Define
  \[
    \mscr S := \set{S\in \mcal S: \text{For all maximum weight $b$-matchings $M$ of $G[S]$, $G[S][M]$ is connected}}.
  \]

  Then $\mscr S$ is a characterization set for the nucleolus of $\Gamma$.
\end{cor}

\begin{pf}
  Fix $S\in \mcal S\setminus \mscr S$.
  Let $M$ be such that $G[S][M]$ has the maximal number of connected components
  among all maximum weight $b$-matchings of $G[S]$.
  By the choice of $S$, $G[S][M]$ must be disconnected.
  Let $T_1, T_2, \dots, T_k$ be the components of $G[S][M]$ for some $k\geq 2$.
  We claim that each $T_i\in \mscr S$.
  Suppose otherwise, that there is some $T_i\notin \mscr S$.
  Then there is a maximum weight $b$-matching $M_i$ in $G[T_i]$ which is disconnected.
  But then $M\setminus E(T_i)\cup M_i$ is a maximum weight $b$-matching in $G[S]$
  with strictly more connected components,
  a contradiction.

  Suppose $x$ is a core allocation.
  Since $x(S) = \sum_{i=1}^k x(T_i)$ and $\nu(S) = \sum_{i=1}^k \nu(T_i)$,
  we have $\sum_{i=1}^k e(S_i, x) = e(S, x)$.
  In particular,
  condition (ii) of \Cref{thm:char-coalitions} is satisfied.
  But all excesses are non-negative as $x$ is a core allocation,
  hence each $e(S_i, x)\leq e(S, x)$
  and condition (i) of \Cref{thm:char-coalitions} is also satisfied.

  The result follows by \Cref{thm:char-coalitions}.
\end{pf}

\begin{lem}\label{lem:pruning}
  Let $\Gamma = (N, \nu)$ be a not necessarily simple weighted $b$-matching game
  with non-empty core.
  Suppose
  \[
    \mscr S := \set{S\in \mcal S: \text{For all maximum weight $b$-matchings $M$ of $G[S]$, $G[S][M]$ is connected}}.
  \]
  is polynomially sized.

  Then the nucleolus of $\Gamma$ is polynomial time computable.
\end{lem}

\begin{pf}
  Apply \Cref{prop:char-coalitions} and \Cref{cor:char-bmatching}.
\end{pf}

\subsection{Simple b-Matching Games}\label{sec:constant 2}
We now present a proof for the first claim of \Cref{thm:b<=2},
which is stated again below for convenience.
Let $G$ be a simple bipartite graph with bipartition $N = A\dot\cup B$
and $k\geq 0$ a universal constant independent of $\card N$.
Let $b\leq 2$ be some node-incidence capacity.
\begin{enumerate}[(i)]
  \item Suppose $b_v = 2$ for all $v\in A$ but $b_v = 2$ for at most $k$ vertices of $B$,
    then the nucleolus of the simple weighted $b$-matching game on $G$ can be computed in polynomial time.
  \item If $b\equiv 2$,
    then the nucleolus of the non-simple weighted $b$-matching game on $G$ can be computed in polynomial time.
\end{enumerate}

\begin{pf}[\Cref{thm:b<=2}(i)]
  By \Cref{lem:pruning},
  it suffices to show that any component of a $b$-matching
  in some arbitrary induced subgraph $G[S]$ has at most $2k+3$ vertices.
  If we show this, then the set
  \[
    \mscr S := \set{S\in \mcal S: \text{For all maximum weight $b$-matchings $M$ of $G[S]$, $G[S][M]$ is connected}}
  \]
  is polynomially sized since it is contained in the subsets of $V(G)$ of size at most $2k+3$.

  Let $C$ be a component of $G[S][M]$ for some $S\sset N$ and maximum weight $b$-matching $M$ of $G[S]$.
  
  If $C$ is a cycle, then exactly half the vertices of $C$ are from $B$ with $b_v = 2$.
  It follows that $\card C\leq 2k$.

  Suppose now that $C$ is some path.
  By deleting at both endpoints and one more vertex,
  we may assume that every other vertex in the path are from $B$ with $b_v = 2$.
  Thus $\card C \leq 2k+3$ as required.
\end{pf}

\subsection{Non-Simple 2-Matching Games}\label{sec:non-simple}
In the case where we allow for edges to be included multiple times in a 2-matching,
we leverage core non-emptiness and the non-existence of odd cycles to compute the nucleolus in polynomial time.

\begin{lem}\label{lem:non-simple 2-matching core non-empty}
  Let $G$ be an arbitrary graph with edge weights $w:E\to \R$.

  The core of the weighted non-simple 2-matching game on $G$ is non-empty.
\end{lem}

Consider the following LP formulation of the maximum weight non-simple $b$-matching from \cite{schrijver2003}.
\begin{align}
  &\max w^Ty &&(P) \label{eq:non-simple 2-matching polytope} \\
  y(\delta(v)) &\leq b_v &&\forall v\in V \\
  y(E[U]) &\leq \floor*{\frac12 b(U)} &&\forall U\sset V, \text{$b(U)$ is odd} \label{eq:b-odd constraints} \\
  y &\geq 0
\end{align}
Observe that for the case $b\equiv 2$,
$b(U)$ is never odd and hence there are no constraints of the form \Cref{eq:b-odd constraints}.

Thus the dual LP of non-simple 2-matching games can be simplified to the following.
\begin{align}
  &\min 2^Tx &&(D) \\
  x_u + x_v &\geq w(uv) &&\forall uv\in E \\
  x &\geq 0
\end{align}

\begin{pf}
  Let $\bar x$ be an optimal solution to $(D)$.
  Since $2\bar x(N) = \nu(N)$ by the integrality of $(P)$,
  $\bar x$ is an allocation.

  Fix a coalition $\varnothing\neq S\subsetneq N$.
  Define $(D_S)$ as the dual to the non-simple 2-matching LP (\Cref{eq:non-simple 2-matching polytope}) on $G[S]$.
  Write $\eval{\bar x}_S$ as the restriction of $\bar x$ to entries indexed by vertices of $S$.

  Observe that $\eval{\bar x}_S$ is feasible in $(D_S)$,
  thus $\nu(S)\leq \bar x(S) = \eval{\bar x}_S(S)$ by weak duality
  and $\bar x(S) - \nu(S) \geq 0$.

  By the arbitrary choice of $S$,
  $\bar x$ is a core allocation
  and consequently, the core is non-empty.
\end{pf}

Notice that since $b\equiv 2$,
we did not need to assume $G$ to be bipartite.

\begin{lem}\label{lem:non-simple 2-matching}
  For any bipartite graph,
  there is a maximum weighted non-simple 2-matching consisting only of parallel edges.
\end{lem}
\begin{pf}
  Let $M$ be a maximum weight non-simple 2-matching in $G$.
  Observe that the components of $G[M]$ are parallel edges,
  even cycles, and paths.
  Moreover, any path contains at least 2 edges,
  or else adding that single edge a second time to our matching can only increase the weight of the matching.

  Let $x$ be the characteristic vector of $M$.
  We argue that all vertex solutions to \Cref{eq:non-simple 2-matching polytope}
  contain no even cycles nor paths containing at least 2 edges.
  Hence an optimal vertex solution has the desired properties.

  \underline{Case I: Even Cycles}
  Suppose $C\sset M$ for some even cycle $C$.
  Enumerate the edges $C: e_1, e_2, e_3, \dots, e_k$
  for some $k\equiv 0\mod 2$.

  For $i=0, 1$, put $M^{(i)} := M\setminus C\cup \set{e_j, e_j: j\equiv i\mod 2}$.
  That is, double up every other edge of $C$.

  Set $x^{(i)}$ to be the characteristic vector of $M^{(i)}$.
  Then $x = \frac12 x^{(0)} + \frac12 x^{(1)}$
  and so $x$ was not a vertex solution.

  \underline{Case II: Paths of Length At Least 2}
  Let $P\sset M$ be a path of length at least 2.
  Enumerate the edges $P: e_1, e_2, \dots, e_k$.
  Similar to the previous case, define for $i=0, 1$
  $M^{(i)} := M\setminus P\cup \set{e_j, e_j: j\equiv i\mod 2}$.

  Set $x^{(i)}$ to be the characteristic vector of $M^{(i)}$.
  Then $x = \frac12 x^{(0)} + \frac12 x^{(1)}$
  and so $x$ is not a vertex solution.
\end{pf}

We are now ready to prove the second claim of \Cref{thm:b<=2}.

\begin{pf}[\Cref{thm:b<=2}(ii)]
  By \Cref{lem:pruning},
  it suffices to show that if $\card S\geq 3$,
  then there is a 2-matching in $G[S]$ with multiple components.

  But this is precisely what we proved in \Cref{lem:non-simple 2-matching},
  concluding the proof.
\end{pf}

Unfortunately, \Cref{lem:non-simple 2-matching} does not hold when the graph is non-bipartite,
even when we restrict ourselves to uniform edge weights.
Indeed, consider the simple triangle.
The maximum cardinality non-simple 2-matching has size 3.
However, when we restrict ourselves to matchings composed of only parallel edges,
the maximum matching we can obtain has cardinality 2.

Similarly, \Cref{lem:non-simple 2-matching} does not in general hold when there are some vertices $v$ where $b_v = 1$.
Consider the path of 3 edges where the endpoints have $b_v = 1$ while the internal vertices have $b_v = 2$.
The maximum cardinality non-simple 2-matching has size 3.
However, if we only allow parallel edges,
the maximum matching we can obtain again has cardinality 2.

On the hopeful side,
we note that a technique used to prove that
the nucleolus of 1-matching games on general graphs with non-empty core can be computed in polynomial time \cite{biro2012}
could possibly extend to non-simple 2-matching games on general graphs.
There,
the authors demonstrated a bijection between the so-called maximum weight half-matchings of a graph
and the maximum weight matchings of a gadget graph.

\section{Conclusion}
We showed that computing the nucleolus
for simple bipartite $b$-matching games when $b\geq 3$ is \NP-hard.
When $b\leq 2$, we described a polynomial time algorithm
for nucleolus computation in simple $b$-matching games on bipartite graphs,
given that one side of the bipartition has a constant number of vertices $v$ where $b_v = 2$.
We also provided a polynomial time algorithm
for computing the nucleolus of non-simple 2-matching games within bipartite graphs.
This is a relaxation of the simple $b$-matching nucleolus problem when $b\equiv 2$.
Our second positive result relies heavily on bipartiteness and the choice of $b\equiv 2$.

As mentioned,
a next step following our work could be to examine the nucleolus computation algorithm for general graphs with non-empty core \cite{biro2012}
and determine whether it can be extended to non-simple 2-matchings in general graphs.
Another future direction could be to provide an efficient algorithm
for simple $b$-matching nucleolus computation in bipartite,
and then in general graphs when $b\leq 2$ or prove that this is \NP-hard.
As the LP-based schemes of Kopelowitz and Maschler have been the basis to our efforts,
it may be of interest to explore combinatorial algorithms to compute the nucleolus.
For instance, Hardwick gives a combinatorial algorithm for some special cases of the 1-matching game \cite{hardwick2017}
and it would be interesting to give a combinatorial algorithm for the general case.

\section{References}
\printbibliography[heading=none]

\section{Appendix}
\listoffigures
\listoftables

\end{document}